\documentclass{kapproc}
\def\be{\begin{equation}}
\def\ee{\end{equation}}
\def\ba{\begin{array}}
\def\ea{\end{array}}
\def\bea{\begin{eqnarray}}
\def\eea{\end{eqnarray}}

\def\tr{{\rm tr}}

\def\chib{{\bar\chi}}
\def\psib{{\bar\psi}}
\def\nn{\nonumber}
\def\wS{S}
\def\wT{T}
\def\sS{{\cal S}}
\def\sT{{\cal T}}
\def\NPB#1#2#3{{Nucl.~Phys.} {\bf{B#1}} (19#2) #3}
\def\PLB#1#2#3{{Phys.~Lett.} {\bf{B#1}} (19#2) #3}
\def\PRD#1#2#3{{Phys.~Rev.} {\bf{D#1}} (19#2) #3}
\def\bi{\bibitem}
\usepackage{latexsym}
\normallatexbib 

\def\np{\mbox{\em {Nucl. Phys.} {\bf B }}}
\def\pl{\mbox{\em {Phys. Lett.} {\bf B }}}

\def\prd{\mbox{\em {Phys. Rev.} {\bf D }}}

\begin{document}

\articletitle{Gaugino Condensation and SUSY Breakdown}

\articlesubtitle{ }
\author{Hans Peter Nilles}

\affil{Physikalisches Institut, Universit\"at Bonn,\\
Nussallee 12, D-53115 Bonn, Germany}
\email{nilles@th.physik.uni-bonn.de}

\begin{abstract}

We review the mechanism of gaugino condensation in the framework of the
$d=10$ heterotic string and its $d=11$ extension of Horava and Witten.
In particular we emphasize the relation between the
gaugino condensate and the flux of the antisymmetric tensor fields of
higher dimensional supergravity.
Its potential role for supersymmetry breakdown and moduli stabilization
is investigated. 

\end{abstract}

\section{Introduction}

The topic of my lectures at the Carg\`ese summer school 2003 was a general
introduction to the
breakdown of supersymmetry in field- and string-theory. In this written up
version I decided to
concentrate on some particular aspects of this mechanism: gaugino condensation
in the
framework of heterotic string- and M-theory. This allows a more detailed
discussion of
supergravity in $d=10$ and $d=11$ dimensions and the reduction to the 
$d=4$ case.

The mechanism of gaugino condensation is believed to play a crucial role for
moduli
stabilization and SUSY breakdown in string theory. Conceived as a mechanism for
hidden
sector supersymmetry breakdown in supergravity extensions of the standard model
of strong and
electroweak interactions \cite{HPN2,2,FGN} 
it found a natural setting in the framework of
the $E_8 \times E_8$
heterotic string theory \cite{Gross:1985fr} (see ref. \cite{DIN,DRSW})
 as well as the M-theory of Horava and Witten \cite{HW} 
(suggested in \cite{H,NOY}).

One of the attractive features of the mechanism is a specific cancellation of
$H-$flux and the
gaugino condensate in the low energy effective potential. This allows a
somewhat controlled
discussion of the vacuum energy at least at the classical level and it is the
direct consequence of
the properties of the higher dimensional supergravity action. It also
emphasizes the importance
of Chern-Simons-terms in $H-$flux of $d=10$ supergravity \cite{DIN2}.

In these lectures we shall not attempt to construct a fully realistic model but
try to explain the
mechanism in its simplest form. In section 2 we will discuss supergravity in
$d=10$ and then
define the supermultiplets relevant for the $d=4$ discussion. The mechanism of
gaugino
condensation is introduced in section 4 followed by a determination 
of the $d=4$ effective action
using the method of reduction and truncation \cite{W2}. 
Section 6 contains some
aspects of the theory
beyond the classical level. In section 7 we give a detailed discussion of the
$d=11$ heterotic M-theory
and the determination of the low energy effective actions. Some aspects 
of the mechanism that
were rather obscure
in the $d=10$ theory (like the cancellation of gaugino bilinears due to
Chern-Simons flux) become
obvious in this generalized picture \cite{NOY2}. 
Section 8 will then focus on the
specific properties of
supersymmetry breakdown at the hidden wall. The last section discusses some
recent
developments that lead to a revived interest of this mechanism during the last
year.

Before we start let me make a technical comment. Traditionally there are two
different ways to
include a gaugino condensate in the effective action. 
The first one \cite{FGN,DIN} uses
explicitly the $F-$terms of the supersymmetry transformation laws. 
These include gaugino bilinears
multiplied by
the derivative of the gauge kinetic function. These gaugino bilinears are then
replaced by the
(field dependent) renormalization group invariant scale 
and included in the standard
fashion in the scalar
potential. This is the procedure which we follow in these lectures. An
alternative method \cite{DRSW,DIN2}
postulates the gaugino bilinear as a new term in an effective superpotential.
The qualitative
results  of the two mechanisms are the same. Some quantitative differences
will be mentioned
where it applies.

\section{Supergravity in $d = 10$ }

We shall here discuss the effective action of superstring theories in 
the supergravity field theory framework.
For the known ($N=1$) superstring theories this is 
$N=1$ supergravity in $d=10$  coupled to pure $E_8 \times E_8$ or $0(32)$  
gauge multiplets. The spectrum of this theory is given by the supergravity 
multiplet $(g_{MN},  \psi _{M \alpha}, B_{MN}, \lambda_\alpha  ,  \phi$) 
where $M, N = 0, \ldots , 9$ are world indices and $\alpha $ is a 
Majorana-Weyl spinor index, as well as the gauge multiplet $(A^{A}_{M}, 
\chi ^{A}_{\alpha })$ where 
$A =1, \ldots, 496$ lables the adjoint representation of $E_8 \times E_8$ 
or $0(32)$. In the Type I theory, these correspond to the massless 
closed (open) string states respectively. The action of such a theory, 
including terms up to two derivatives, 
is unique and given by \cite{CM}:
\begin{eqnarray}
\lefteqn{e^{-1}_{10}  {\cal L}= - \frac{1}{2} R - \frac{i}{2} \bar{\psi}^{M} 
\Gamma^{MNP}
 D_N ( \omega )\psi_P + 
\frac{9}{16} \left( \frac{\partial_M \varphi}{\varphi}\right)^2 +} \nonumber\\
&& + \frac{3}{4}\varphi^{- 3/2} H_{MNP} H^{MNP} + \frac{i}{2} \bar{\lambda} 
\Gamma^M 
D_M  \lambda + \frac{3 \sqrt{2}}{8} \bar{\psi}_M 
\left( \frac{\Gamma^P \partial_P \varphi}{\varphi}\right) \Gamma^M \lambda
\nonumber\\
&& - \frac{\sqrt{2}}{16} \varphi^{- 3/4} H_{MNP} \left( 
i \bar{\psi}_Q \Gamma^{QMNPR} \psi_R + 6 i \bar{\psi}^M \Gamma^N \psi^P + 
\right.
\nonumber\\
&& \left.+ \sqrt{2} \bar{\psi}_Q \Gamma^{MNP} \Gamma^Q \lambda - i \bar{\chi} 
\Gamma^{MNP} \chi \right) - 
\nonumber\\
&& - \frac{1}{4} \varphi^{-3/4} F_{MN} F^{MN} + \frac{i}{2} \bar{\chi} 
\Gamma^M D_M (\omega ) \chi - 
\nonumber\\
&& - \frac{i}{4} \varphi^{- 3/8} \left( \bar{\chi} \Gamma^M \Gamma^{NP} 
F_{NP} \right)
\left( \psi_M + \frac{i \sqrt{2}}{12} \Gamma_M \lambda \right)
\nonumber\\
&& + \; \mbox{four fermion interactions}
\end{eqnarray}
where $\Gamma$ denote Dirac matrices in $d=10$ and
\begin{equation}
F^{A}_{MN} = \frac{1}{2} \partial_{ [ M} A^A_{N]} + f^{ABC} A^B_M A^C_N
\end{equation}
(written for short as $F= dA + A^2$)  denotes the gauge field strength. 
Supersymmetry requires the field strength $H_{MNP}$
of the antisymmetric tensor field $B_{MN}$ not just to be the curl of B, but
\begin{equation}
H^A_{MNP} = \partial_{[M} B_{NP]} + \omega^{YM}_{MNP}
\end{equation}
where the Chern-Simons term is given by
\begin{equation}
\omega^{YM} = Tr \left( AF - \frac{2}{3} A^3 \right)
\end{equation}
i.e., $B_{NP}$ has to transform non-trivially under the $E_8 \times E_8$ 
[or $0(32)]$ gauge transformations. This theory as it stands has 
gravitational anomalies and is too naive an approximation to the 
anomaly-free superstring theory. The absence of anomalies 
requires an additional term to (3)\cite{GS}:
\begin{equation}
H = d B + \omega ^{YM} - \omega ^L
\end{equation}
with
\begin{equation}
\omega ^L = Tr ( \omega  R - \frac{2}{3} w^3 )
\end{equation}
where $\omega ^{ab}_M$ is the spin connection. $\omega$ contains a 
derivative, thus $\omega ^L$ contains three and appears squared in 
the action. This term is purely bosonic and for a supersymmetric 
action requires additional terms which up now are only partially 
known. The action in (1) thus requires further terms in order to 
be an adequate low-energy limit of string theory. The action (1) 
was derived by truncating all heavy string states. For a better 
approximation they should be integrated out, leaving a low-energy 
theory with higher derivatives and terms in a higher order in 
$ \alpha'$ (the slope parameter). These terms appear in what is 
usually called  ''$\sigma$ -model perturbation theory'', not to 
be confused with the string loop expansion, which, at least in 
the heterotic case, is an expansion in $g$, the gauge coupling 
constant. This expansion in powers of 
$ \alpha '$
is classical at the string level. There might also be world-sheet 
non-perturbative effects that play a role at this classical level. 
Looking at (1), one might wonder what $g$ (the gauge coupling constant) 
is. Observe that the gauge fields have non-minimal gauge kinetic terms. 
Here $g$ is not an input parameter, but $g$ will be determined dynamically.
\begin{equation}
\frac{1}{g^2} = < \varphi ^{- 3/4} >
\end{equation}
consistent with the expectations in string theory. We have to be 
aware of the fact that the coupling constant as determined by this 
naive approximation might be different from that determined by  
string theory. 
This approximation 
is probably only useful in defining the important interactions at 
low energies. In order to ask more fundamental questions, like the 
determination of the fundamental coupling constants, the approximation
 probably has to be improved. This can already be seen when we discuss 
compactification. One possible way is to compactify on a six-torus $T^6$ , 
leading to $N=4$ supergravity in $d=4$, which does not resemble known  
$d=4$ phenomenology. One might therefore ask the question for more 
non-trivial compactifications (still postponing the question of 
why these should be more likely than the trivial ones). Defining 
$\phi = (3/4 ) \log  \varphi$ and neglecting fermionic terms, 
the equation of motion for $\phi$ is:
\begin{equation}
\Box \phi = \exp ( - \phi ) \left[ F^2_{MN} + \exp ( - \phi ) H^2_{MNP} \right]
\end{equation}
Integrating $\Box \phi$ over a compact manifold without boundary 
leads to a vanishing result. The right-hand side is positive definite 
and therefore has to vanish. This implies trivial compactification unless 
$\phi \to \infty$, which is outside the validity of our approximation. 
The addition of $\omega^L$ 
 in $H$ does not change the situation, but this term requires 
supersymmetric completion which necessitates the presence of $R^2$ terms. 
They actually appear in the Euler combination
\begin{equation}
- \exp ( - \phi ) \left[ R^2_{MNPQ} - 4 R^2_{MN} + R^2 \right]
\end{equation}
on the right-hand side of (8), ensuring the absence of ghosts. 
With these terms from the $\alpha^\prime$ expansion, non-trivial 
compactification is possible: $R^2$ can be compensated by $F^2$, 
and this implies a breakdown of gauge symmetries in the presence 
of compactification \cite{Cetal}. 
Notice, however, that the scale of compactification 
is not yet fixed. There exists an independent argument confirming 
this result. For the $H$ field to be well defined, the integral 
of the curl of $H$ over a compact manifold without boundary should vanish:
\begin{equation}
\int\limits_{C_4} dH = \int\limits_{C_4}  \left[ Tr F \wedge F - Tr R \wedge R
 \right] = 0
\end{equation}
leading to a compensation of $F$ and $R$ in extra dimensions. 
These results are very encouraging. If 
$E \times E_8 $ or $0(32)$ were to remain unbroken in $d= 4$, 
they would not be able to lead to chiral fermions. The discussed 
constraints involve integrated quantities and could have various 
solutions. Only the simplest possibility -- a vanishing integrand -- 
can be studied easily \cite{W2}. 
It implies a direct identification of $F$ and $R$. 
The spin connection 
$\omega^{ab}_m (m = 4, \ldots , 9 \; a,b = 1. \ldots , 6 )$ 
can be viewed as a gauge field of an $0(6)$ subgroup of the Lorentz 
group $0(9, 1)$, identified with $A^{A}_m$ in an $0(6)$ subgroup of 
$E_8 \times E_8$ or $0(32)$ in order to fulfil the constraints. 
The question of a remaining supersymmetry in 
$d=4$ is related to the holonomy group of the compact manifold, 
which in turn is a subgroup of $0(6)$. I shall not explain this 
relation here in detail, but just give a heuristic argument. 
The gravitino $\psi^\alpha_M$ transforms like a 4 of $0(6)$. 
$N=1$ supersymmetry will be present in $d=4$ if the decomposition 
of the 4 with respect to the holonomy group contains exactly one 
singlet. If there are more singlets, one will have extended 
supersymmetries, e.g., in the case of the torus the holonomy 
group is trivial and $4 = 1 + 1 + 1+ 1$, resulting in $N=4$ 
supersymmetry. The simplest choice for $N=1$ is to have 
$SU(3)$ holonomy, which leads to $4 = 1 + 3$ and $6 = 3 + \bar{3}$, 
and is used in the Calabi-Yau approach. But there are certainly 
more possibilities, even with discrete subgroups of $SU(3)$ corresponding 
to certain orbifolds. For simplicity, I shall here assume $SU(3)$ 
holonomy. With this identification of $\omega$ and $A$ at least 
one $SU(3)$ subgroup of $0(32)$ or $E_8 \times E_8$ will break 
down during compactification. In the case of $0(32)$, this will 
lead to $0(26) \times U(1)$ with possible zero modes in the 
decomposition of the adjoint of $0(32)$, giving exclusively real 
representations of $0(32)$. Based on this argument, one usually 
concludes that $0(32)$ will not lead to a phenomenologically successful 
model, although not all possibilities have yet been studied. 
The situation in the case of $E_8 \times E_8$ looks better. 
A decomposition of the adjoint of $E_8$ with respect to 
$E_6 \times SU(3)$ leads to 
$248 = (78, 1) + (27, 3) + (\overline{27}, \bar{3} + (1, 8)$ 
and contains chiral representations. Moreover, $E_6$ is one 
of the more successful candidates for a grand unified 
gauge group with a family of quarks and leptons in 27, 
the number of these zero modes being defined by topolocigal 
properties of the compact manifold. Here is then a common 
starting point for the construction of ''superstring-inspired models''.


\section{Towards $d=4$}

We have first to discuss the possible zero modes. Let us define indices 
$M= (  \mu , m )$  $( \mu = 0, \ldots , 3 ; \; m = 4 , \ldots , 9)$ 
and start with the metric
\begin{equation}
g_{MN} = \left( \begin{array}{c|c}
g^{- 1/2}_6 \hat{g}_{\mu \nu}& \phantom{g^{- 1/2}_6 \hat{g}_{\mu \nu}}\\
\cline{1-2} & g_{mn}
\end{array}
\right)
\end{equation}
where $g_6 = \det g_{mn}$ is used to redefine $g_{\mu \nu}$ 
in order to have usual kinetic terms for the graviton. 
The integral over extra dimensions
\begin{equation}
\int\limits d^6 y \sqrt{- g_6} = R^6_c \sim \frac{1}{M^6_c}
\end{equation}
defines the average radius of compactification. Defining 
$g_{mn} = \exp (\sigma ) \hat{g}_{mn}$, one can then 
normalize $\int\limits d^ 6 \sqrt{- \hat{g}_6} = M^{-6}_P$ and 
$\exp (\sigma )$ defines the radius of compactification in 
units of the Planck length. Depending on the topological properties 
of the manifold, $g_{mn}$ gives rise to zero modes that are scalars 
in $d=4$ (we will not discuss off-diagonal terms in $g_{MN}$  
like $g_{\mu m}$ that give rise to gauge bosons depending 
on the isometries of the manifold). $g_{mn}$ corresponds to 
a symmetric tensor of $0(6)$ with respect to the $SU(3)$ 
subgroup discussed earlier; we have $21 = 1 + 8 + 6 + \bar{6}$. 
With the notation $m = ( i , \bar{j})$, the latter correspond 
to modes of 
$g_{i \bar{j}}, g_{ij}, g_{\bar{i}\bar{j}}$ , while $\sigma$ 
is the singlet.

Turning to the gravitino $\psi^\alpha_M$, we can view $\alpha$ as 
an eight-dimensional index which transforms as a $4$ of $0(6)$ and 
a Weyl spinor of $0(3, 1)$. $\psi^\alpha_\mu$ corresponds to spin-3/2 
particles in $d= 4$ with $N_{max} = 4$ as already discussed. $\psi^\alpha_m$ 
can give rise to spin-$\frac{1}{2}$ zero modes. To obtain canonical 
kinetic terms for the gravitino, as in the case of the metric, a rescaling
\begin{equation}
\tilde{\psi}_\mu = \exp ( - 3 \sigma /4 ) \psi_\mu
\end{equation}
is required.

The antisymmetric tensor field $B_{MN}$ could give rise to 
$B_{\mu \nu}, B_{m \nu}$ and $B_{mn}$ (corresponding to the 
Betti numbers $b_0 , b_1$ and $b_2$). A zero mode from 
$B_{\mu \nu}$ corresponds to one pseudoscalar degree of 
freedom $\theta$ defined through a duality transformation
\begin{equation}
H_{\mu \nu\varrho} \epsilon^{\mu\nu\varrho\sigma} = \varphi^{3/2} 
exp (- 6 \sigma) \partial^\sigma \theta + \ldots 
\end{equation}
$B_{m\nu}$ could give rise to extra gauge bosons which 
(although possibly interesting) we shall not discuss here. 
$B_{mn}$ will again correspond to pseudoscalars in $d = 4$. 
A decomposition with respect to $SU(3)$ gives $15 = 1 + 3 + \bar{3} + 8$ 
with the singlet corresponding to the "trace" 
$\eta = \epsilon^{mn} B_{mn}$ and $B_{\bar{i} \bar{j}}$, and 
$B_{ij}$ and $B_{i\bar{j}}$ corresponding to $3, \bar{3}$, 
and $8$ respectively. All these modes appear in the action 
only through the field strength $H$ implying derivative couplings, 
i.e., they show axion-like behaviour. From the $\lambda, \phi$ 
members of the supergravity multiplet, we expect additional 
spin $- \frac{1}{2} (0)$ particles in $d = 4$. 

The discussion of the zero modes of $A^A_M$ involves some complication 
because of the identification of $\omega^{ab}_m$ and $A^A_m$ in an 
$SU(3)$ subgroup. $A^A_\mu$ will of course, give rise to gauge 
bosons in the adjoint representations of the unbroken gauge group, 
e.g., $A = 1$, $\ldots$, $78$ for $E_6$. $A^A_m$ will give rise 
to scalars in $d = 4$, and we are mostly interested in those 
transforming as $27$ (or $\overline{27}$) under $E_6$. Let us 
therefore write $A = (a,i)$ or ($\bar{a},\bar{i})$  $a = 1, \cdots, 27$. 
The states $C^b = A^{b, i}_{\bar{i}}$ and $B^{\bar{b}} = A^{\bar{b}, 
\bar{i}}_i$ then transform as $27, \overline{27}$ with respect to 
$E_6$ and are singlets under the diagonal subgroup $SU(3)$ of the p
roduct of $SU(3) \subset 0(6)$  and $SU(3) \subset E_8$. These 
bosons will have supersymmetric partners from the zero modes 
of $\chi^A_\alpha$. The number of the possible zero models is 
of course entirely defined by the topological properties of 
the manifold under consideration.

We can now have a first look at the possible interactions of 
these zero models in $d = 4$ starting from the $d = 10$ action 
given in (1). Of course, in general we expect here not only the 
influence of topological properties, but also the explicit form 
of the metric of the compact manifold will become important. 
Nonetheless we will be able to obtain some non-trivial results 
that are rather independent of the special form of the metric. 
We will do that exclusively in the framework of $N = 1$ supergravity 
in $d = 4$, firstly because of the reasons given in Section 2, and 
secondly because this theory is simpler than the non-supersymmetric case.

$N = 1$ supergravity in $d = 4$ (with action including terms up 
to two derivatives \cite{Cremmer}) is defined 
through two functions of the chiral superfields $ \phi_i$. The first 
is an analytic function $f (\phi_i)$ defining the gauge kinetic 
terms $f (\phi_i) W^\alpha W_\alpha$. In a component language, 
$f$ appears in many places, but it can be extracted most efficiently from
\begin{equation}
\makebox{Ref} (\varphi_i) F_{\mu\nu} F^{\mu\nu} +\makebox{Imf} 
(\varphi_i) \epsilon_{\mu\nu\varrho\sigma} F^{\mu\nu} F^{\varrho\sigma}
\end{equation}
where $\varphi_i$ denotes the (complex) scalar component of $\phi_i$. 
The second is the so-called K\"ahler potential
\begin{equation}
{G} (\phi_i, \phi^\ast_i) = K (\phi_i, \phi^\ast_i) + \makebox{log} 
| W (\phi_i) |^2
\end{equation}
Unlike $f$, $G$ is not analytic and contains the left-handed 
chiral superfields along with their complex conjugates. 
The second term in (16) contains the analytic function 
$W(\phi_i)$: the superpotential. The action in component 
form usually contains $G$ in complicated form; the scalar 
kinetic terms, e.g., are
\begin{equation}
{G}^j_i ( \partial_\mu \varphi^i) (\partial^\mu \varphi^\ast_j)
 ;  \; \; {G}^j_i \equiv  
\frac{\partial^2 {G}}{\partial \varphi^i \partial \varphi^\ast_j}
\end{equation}
whereas the scalar potential is given by
\begin{equation}
V = {\exp} ({G}) [{G}_k ( {G}^{-1}) ^k_l {G}^l - 3]
\end{equation}
which makes ist difficult to extract $G$ once an action is given in 
component form. There is only one term which allows a rather simple 
identification of $G$, and this is a term involving the gravitino
\begin{equation}
e_4 \exp \left( {G} /2 \right) \bar{\psi}_\mu \gamma^{\mu \nu} 
\gamma_5 \psi_\nu
\end{equation}
which will later be used extensively after the correct redefinitions 
of the gravitino in $d = 4$ have been performed.
Let us now consider the action in $d = 10$ in order to learn something
 about the possible action in $d = 4$. We start with the gauge kinetic term
\begin{equation}
e_{10} \varphi^{- 3/4} F_{MN} F^{MN}
\end{equation}
Since we are interested in the $F^2_{\mu\nu}$ part, we write 
\begin{equation}
e_4 e_6 \varphi^{- 3/4} F_{\mu\nu} F_{\varrho\sigma} g^{\mu\varrho} 
g^{\nu\sigma}
\end{equation}
where, with the definitions given earlier, we would like to extract $f$ from
\begin{equation}
\hat{e}_4 \makebox{Ref} \, F_{\mu\nu} F^{\mu\nu}
\end{equation}
with $\hat{e}_4 = (det  \; \hat{g}_{\mu\nu})^{\frac{1}{2}} = {\exp} 
(6 \sigma) e_4$, and indices are contracted with the "hatted" metric. 
Integrating the extra six dimensions with the normalization given 
in (13) using $M_P \equiv 1$, we obtain 
\begin{equation}
ReS \equiv Ref = \varphi^{-3/4} {\exp} (3 \sigma)
\end{equation}
as the real part of the scalar component of a chiral superfield 
denoted by $S$. This is a rather amazing result. Remember that at 
no point in the derivation did we have to know something about the 
metric of the compact six-dimensional space, so this constitutes a 
rather model-independent result. Observe that $f$ is usually 
non-trivial, that its vacuum expectation value (vev) will determine 
the gauge coupling constant, and that the couplings of 
$E_8$ (or $E_6$) and $E_8'$ coincide.

Let us now discuss the imaginary part of $f$, to be extracted 
from $F_{\mu \nu} F_{\rho \sigma} 
\varepsilon^{\mu \nu \rho \sigma}$. The relevant degree of freedom 
comes from $B_{\mu \nu}$ as discussed earlier. $B_{\mu\nu}$ 
couples only through its field strength $H_{\mu \nu \rho}$ and 
has therefore only derivative couplings. Taking the relevant 
terms in the $d = 10$ action and integrating the extra dimensions, we obtain
\begin{equation}
\varphi^{-3/2} \exp ( 6 \sigma) H_{\mu \nu \varrho} H^{\mu \nu \varrho} 
+ H_{\mu \nu \varrho} O^{\mu \nu \varrho}
\end{equation}
where $O^{\mu \nu \varrho}$ contains fermion bilinears. $H$ has to 
satisfy a constraint (neglecting $R^2$-terms for the moment)
\begin{equation}
\partial_{ [\mu}  H_{\nu \varrho \sigma ] } = - Tr F_{ [ \mu \nu} 
F_{\varrho \sigma]}
\end{equation}
which we take into account by adding a Lagrange multiplier
\begin{equation}
\theta \epsilon^{\mu \nu \varrho \sigma} \left( \partial_\mu 
H_{\nu \varrho \sigma} + Tr F_{\mu \nu} F_{\varrho \sigma} \right)
\end{equation}
Next we eliminate $H$ via the equations of motion and arrive 
at an action containing the terms
\begin{equation}
\varphi^{3/2} \exp ( - 6 \sigma ) ( \partial_\mu \theta)^2 + 
\theta \epsilon^{\mu \nu \varrho \sigma}
Tr ( F_{\mu \nu} F_{\varrho \sigma})
\end{equation}
which tells us that $Imf = \theta$, and for the scalar 
component of $S$ we obtain
\begin{equation}
S = \varphi^{- 3/4} \exp ( + 3 \sigma ) + i \theta
\end{equation}
as a mixture of $g_{MN}$ and $B_{MN}$ zero modes. The   
partner is a combination of $\psi_m$ and $\lambda$ zero modes 
which we will not discuss here in detail. Observe that $\theta$ couples 
only with derivatives except for the last term in (27), and that the 
$d= 4$ action, has a Peccei-Quinn-like symmetry under shifts of 
$\theta$ by a real constant, thus $\theta$ couples like an axion. 
Let me stress again that all these statements about the action 
and the form of (28) are model-independent and could be derived 
without explicit knowledge of the metric.

Unfortunately, the situation changes once we try to extract the
 K\"ahler potential. As already indicated, the term to investigate 
is the $d=4$ ''gravitino mass term'' (19). The extraction of this
 term is rather complicated due to several redefinitions of the 
gravitino field. A general form has been given in 
\cite{DIN2}), and we 
will not repeat the derivation here. Many of the terms appearing 
there depend explicitly on the metric and spin-connection of the 
six-dimensional compact space. A model-independent statement can 
only be made about the structure of the superpotential, because it 
is an analytic function in the chiral superfields. Symbolically 
the ''gravitino mass term'' is obtained as 
\begin{equation}
\exp ( G/2 ) = \varphi^{-3/4} \exp (- 3 \sigma ) \Gamma^{mnp} H_{mnp}
\end{equation}
and from (16) we can try to read off the superpotential. $W(\phi_i)$ 
is defined to be an analytic function in the chiral superfields and 
should not contain derivatives. A first inspection of (29) therefore 
suggests that a possible candidate for a superpotential is the $A^3$ 
term contained in the Yang-Mills Chern-Simons term (4) included in $H$. 
This then gives rise to a trilinear superpotential involving the $C$ and 
$B$ fields defined earlier. At the moment it is not clear whether 
these are the only possible terms in the superpotential, although 
at the classical level
this seems to be 
the complete expression. Observe that, for example, the superfield 
$S$ as defined in (28) cannot appear in the superpotential, since 
its pseudoscalar component has only derivative couplings. We will 
come back to these points later. In any case, a more detailed discussion 
of the K\"ahler potential requires more information (or approximations) 
about the $d=6$ metric. Before we tackle this topic, let me first 
present a discussion about supersymmetry breakdown in $d=4$.

\section{Gaugino condensation and supersymmetry breakdown}

$N=1$ supergravity in $d=4$ still needs the incorporation of supersymmetry 
breakdown at a scale small compared to the Planck mass. For the 
phenomenological reasons mentioned earlier, this should appear in 
a hidden sector only coupled gravitationally to the observable sector. 
Some superstring models contain such a hidden sector, e.g. the 
sector that contains the particles transforming non-trivially under 
the second $E_8$. Notice that the observable sector 
(for definiteness called the $E_6$ sector) only couples gravitationally 
to the $E_8^\prime$ sector (there are no particles that 
transform non-trivially both under $E_6$ and $E_8^\prime$). Moreover, 
the $E_8^\prime$ sector contains a $d= 10$ pure super-Yang-Mills 
multiplet, suggesting a possible breakdown of supersymmetry via 
gaugino condensates. This breakdown has already been discussed 
in the framework of supergravity models, both at the level of an 
effective Lagrangian \cite{HPN2}
and at the level of the complete classical 
action \cite{FGN}. 
Assume asymptotically-free gauge interactions (here 
$E_8^\prime$ or a subgroup thereof) with a scale
\begin{equation}
\Lambda = \mu \exp \left( - 1/b_0 g^2 (\mu ) \right)
\label{eq:lambda}
\end{equation}
which is renormalization-group invariant at the level of the 
one-loop $\beta$-function. In analogy to QCD, which leads 
to $q\bar{q}$ condensates, we will here assume that the 
gauge fermions condense at a scale
\begin{equation}
< \chi \chi > = \Lambda^3
\end{equation}
As long as $\Lambda$ is small compared to $M_p$, we assume that 
gravity will not qualitatively disturb this dynamical mechanism. 
The question whether such a condensate breaks supersymmetry can 
be studied by investigating the supersymmetry transformation laws 
of the fermionic fields of the theory. The non-derivative terms in 
these transformations will give us the auxiliary fields that serve as 
order parameters for supersymmtry breakdown. The relevant objects 
here are the auxiliary fields of the chiral superfields
\begin{equation}
F_k = \exp (G/2 ) G_k - \frac{1}{4} f_k (\chi \chi ) + \ldots
\end{equation}
where $f$ is the gauge kinetic function discussed earlier and 
$f_k$ is its derivative with respect to $\phi_k$. A necessary 
condition for the breakdown of supersymmetry via gaugino condensates 
is therefore a non-trivial $f$-function. This condition is fulfilled 
in the framework of superstring-inspired models \cite{DIN} 
\cite{DRSW}, 
since we have seen 
in the last section that $f= S$ in a rather model-independent way. 
Whether this is also sufficient for the breakdown of supersymmetry 
can only be checked by minimizing the potential
\begin{equation}
V = F_k \left( G^{-1} \right)^k_\ell F^\ell - 3 \exp (G) 
\end{equation}
since the different terms in (32) might cancel at the minimum. But 
let us for the moment assume that only the second term in (32) 
receives a $vev$. Since $f_s = 1$ in units of $M_p$, we find 
a supersymmetry breakdown scale
\begin{equation}
< F_s > = M^2_s \approx \Lambda^3 /M_p
\end{equation}
and a scale of $\Lambda \sim 10^{13}$ GeV would lead to a 
gravitino mass in the TeV range. Once we understand why 
$\Lambda$ is five orders of magnitude smaller than $M_p$, 
we shall understand why $m_{3/2}/M_p \sim 10^{-15}$. $\Lambda$ now 
depends on the $E_8^\prime$ gauge coupling and the spectrum of low-energy 
modes. Identifying $g_6$ with $g_8$ would in many circumstances 
lead to too large a value for $\Lambda$, and one might speculate 
that $E_8^\prime$ should break during compactification. We shall, 
however, see later that the equality of $g_6$ and $g_8$ seems to 
be only an artifact of the classical approximation, which is not 
true in the full theory. Thus the shadow $E_8^\prime$
(or a subgroup thereof) sector of 
the superstring takes the role of the hidden sector of supergravity 
models and might explain the smallness of $m_{3/2}$ compared to 
$M_p$. But how does this breakdown of SUSY in the hidden sector 
influence the observable sector? In general, we would expect 
gaugino masses $(m_{1/2})$, scalar masses $(m_{0})$ and the 
trilinear couplings $(Am)$ to be of the order of magnitude of 
$m_{3/2}$. A naive inspection shows that this might also be 
true here. Gaugino masses in the observables sector are in 
general given by
\begin{equation}
m_{1/2} = f_k \left( G^{-1} \right)^k_\ell F^\ell
\end{equation}
where $f$ is the gauge kinetic function of the observable sector. 
With $F^\ell = (1/4 ) f^\ell < \chi \chi>$ we would therefore 
obtain $m_{1/2} \sim m_{3/2}$. In the same way we would obtain under 
these circumstances the soft
trilinear couplings $A\sim 1$ and scalar masses of order $m_{3/2}$. 
To make a quatitative statement about the soft parameters
we need a better understanding 
of the K\"ahler potential, a question which we want to discuss
in the next section. We need this in order to study the explicit
form of the effective potential and finds its minimum.
This then also has to determine the exact
value of $\Lambda$ in (\ref{eq:lambda}) which depends on the
coupling constant and is fixed only after the value of $g$ is
known. 


\section{Reduction and Truncation}

A first approximation for $G$ (that might simulate an orbifold 
approximation of interest in this context) is obtained through 
reduction and truncation \cite{w1}.
One first compactifies the $d= 10$ theory  
on a six-torus $T^6$. The resulting theory is $N= 4$ supersymmetric 
in $d=4$. From this theory one truncates unwanted states, to obtain 
an $N= 1$ theory. From the gauge singlet sector one keeps only 
those states that transform as singlets under an $SU(3)$ $\subset 0(6)$ 
of the Lorentz group. Since $\psi^\alpha_\mu$ transforms as a $4$ of 
$0(6)$ and thus as 1 + 3 under $SU(3)$, we remain with one gravitino. 
As already explained in Section 3, there are only a few gauge singlets 
that survive this truncation. For the bosonic modes we have 
$\varphi$, $\sigma$ from the metric as well as $\theta$ and 
$\eta$ from the antisymmetric tensor. For the gauge non-singlet 
fields one has to remember the identification of spin-connection 
and gauge fields. Here one keeps those states which are singlets 
under the diagonal subgroup of the product of $SU(3) \subset 0(6)$
 and $SU(3) \subset E_8$. This leaves us with one 27 of $E_6$ 
in this case, corresponding to $C^b = A^{b, i}_{\bar i} $; 
($b = 1, \ldots , 27,$ cf. Section 3). With this well-defined 
procedure based on simple reduction on $T^6$, the component 
Lagrangian in $d=4$ can be deduced. From this we can 
immediately read off $f= S$ and $W = d_{abc} C^a C^b C^c$, 
which should not be surprising. Moreover, from the ''gravitino mass term'' 
formula (29) one obtains
\begin{equation}
{G} = \log \left( e^{- 6 \sigma} \varphi^{- 3/2} \right) + \log |W|^2
\end{equation}
The components $\varphi$ and $\sigma$ should correxpond to 
lowest components of chiral superfields. One combination 
$S= \varphi^{- 3/4} \exp (3 \sigma ) + i \theta$ has already 
been defined earlier. To define the other combination, 
the information from (36) is not enough. The charged fields $C$ 
do not yet appear in the first term of $G$  in (36) and the 
correct definition of the superfields has yet to be found. 
This can be done, for example, by using the scalar kinetic terms. 
It leads to a second superfield in which $\varphi , \sigma$ and 
the $C$-modes mix
\begin{equation}
T = \exp ( \sigma ) \varphi^{3/4} + | C_a |^2 + i \eta
\end{equation}
where $\eta$ is the mode from $\epsilon^{mn} B_{mn}$ as discussed 
earlier, and the K\"ahler potential from (36) thus reads
\begin{equation}
{G} = - \log \left( S + S^* \right) - 3 \log \left( T + T^* - 
2 | C |^2 \right) + \log | W | ^2
\end{equation}
a form already previously mentioned in the framework of supergravity 
models. The scalar potential derived from this $G$-function has 
some remarkable properties
\begin{equation}
V = \frac{1}{16 st^3_c} \left[ | W | ^2 + \frac{t_c}{3} | W^\prime 
|^2 \right] + D^2 - \mbox{terms}
\end{equation}
where $s = Re  S$ and $t_c = Re T - | C_a |^2 = t - | C_a |^2$ and 
$W^\prime$ is the derivative of $W$ with respect to the $C$-field. 
The potential is positive definite ($ t_c > 0$ is required by the 
kinetic terms ) and has a minimum with vanishing vacuum energy $V = 0$. 
This minimum is obtained at $W = W^\prime = 0$ independent of the 
values of $s$ and $t$. This implies that at this level the gauge 
coupling constant and the radius of compactification is not yet fixed. 
The theory has classical symmetries which allow shifts of the values 
of $s$ and $t$, as well as Peccei-Quinn symmetries corresponding 
to shifts in $\theta$ and $\eta$. This, of course, makes the use 
of this approximation as an effective low-energy limit of the 
superstring very problematic. 
Certain crucial parameters, like the value of the gauge coupling 
constant and the scale of compactification, which we believe to 
be dynamically determined in the full string theory, are not yet 
fixed. To determine these quantities we would need information
beyond the truncated theory. 

This remains a relevant question when we discuss the effective
potential  in the presence of a gaugino condensate.
Since the gauge 
coupling constant is not determined, $\Lambda$ in (30) is 
also unknown. Using (33) and (38), we get for the potential 
\begin{equation} 
V = \frac{1}{16st^{3}_{c}} \left[ |W-2(st_{c})^{3/2}(\chi\chi)
|^{2}+\frac{t_c}{3} |W^\prime|^{2} \right]
\end{equation}
where $(\chi\chi)$ depends on $g^{2}$ through $\exp(-S/b_{0})$. 
The potential is still positive definite and has a minimum at $V = 0$ 
which is still degenerate. Now the minimum need not necessarily 
imply $W = W^\prime = 0$, but we could have a non-trivial vev 
of $W$. Given fixed $<W> \: \neq \:  0$ by some yet unknown mechanism, 
the value of the gauge coupling constant would be fixed. 
The most natural candidate for such a mechanism would be a
nontrivial vev (so-called flux) of the antisymmetric tensor field
$H$ as defined in (5). At first sight one might have conjectured that 
the appropriate flux would originate from $dB$, but it was soon realized 
that $<dB>$ is quantized in units of the Planck scale \cite{RW}. Thus
$<dB>$ should vanish in order to allow for a value of
$<\chi\chi>$ that is small enough to give a reasonable value for the
gravitino mass after supersymmetry breakdown. One therefore
concluded that it is the flux of the Chern-Simons terms in (5)
which is responsible for the non-zero $H-$flux \cite{DIN2}. For 
these terms the quantization argument of ref. \cite{RW} does not 
apply and acceptable values for $<\chi\chi>$ can be obtained.
We shall come back to this question in more detail in the framework of the
heterotic M-theory. There it will become obvious from theoretical
arguments that ist is not $<dB>$ but the flux from the Chern-Simons
terms that compensates the contribution of the gaugino condensate
in the effective potential.

In order to minimize the potential, the theory slides to a 
coupling constant which, through (30), gives a value of the 
condensate that exactly cancels the contribution of $W$. In 
other words, this means that the dilaton $S$ slides to a value 
that cancels the vacuum energy in the same way as an axion slides 
to cancel a possible $\theta$-parameter of a gauge theory 
[observe that $\exp(-S/b_{0})$ contains both $s$ and $\theta$]. 
Although we do not yet understand the magnitude of supersymmetry 
breakdown, this mechanism to ensure $E_{\rm vacuum} = 0 $ after 
$SUSY$ breakdown 
appears very attractive. 
We shall still need to convince ourselves 
that supersymmetry is actually broken, since in (40) a certain 
cancellation of $<W>$ and $<\chi\chi>$ appears. In fact it tells 
us that the auxiliary field $F_S$  of the $S$-superfield vanishes 
in the vacuum. Nonetheless, here $F_T$ requires a non-vanishing 
vev once $<W> \: \neq \: 0$, and supersymmetry is broken
\begin{equation}
F_T=exp (G/2) G_T \: \neq \: 0.
\end{equation}

In a next step we have to analyze how the breakdown of $SUSY$ is felt in 
the observable sector, and this is, of course, model dependent. 
Gaugino masses, for example, 
are given by $m_{1/2} = f_k (G^{-1})^k_\ell F^\ell $, 
and only $f_S$ is different from zero. In the case at hand 
we therefore 
obtain $m_{1/2} = 0$. In fact, the same is also true for the
scalar masses. This is an artifact of the special model
(a so-called no scale model) and would need further discussions.
A first question concerns the stability of this result
in perturbation theory which we shall investigate in the next
section.

Before we do this let us mention a different way to include the 
gaugino condensate in the low-energy effective potential
\cite{DRSW}. Instead of including $(\chi\chi)$ in $F_k$ directly,
as in (32), one might postulate a new contribution to the
superpotential proportional to
$(\chi\chi)\sim \exp(-S/b_0)$. This leads to a potential very
similar although not identical to the one given in (40) 
(for details see \cite{DRSW,DIN2}). 


\section{Beyond the classical level}

Vanishing values of the soft parameters in the observable sector
might be a result of the symmetries of the theory, and if yes,
whether these symmetries hold to all orders in the perturbative
loop expansion.
In the heterotic string this 
loop expansion is governed by the coupling constant $g$, which 
in turn is defined through a $vev$ of the dilation field. This 
will allow us to construct a definite loop expansion in the 
dilation field and still give us restrictions on how the 
classical symmetries are broken by loop effects. But before we 
discuss the loop expansion in more general terms, let us examine 
some aspects at the one-loop level.
We can do that because of the mechanism of anomaly cancellation 
in the $d = 10$ field theory. Green and Schwarz have observed 
that the cancellation of anomalies 
\cite{GS} requires certain new local 
counterterms with definite finite coefficients in the one-loop 
effective action to cancel the gauge non-invariance of present 
non-local terms. In general, such terms appear with infinite 
coefficients, but the possible symmetry of the effective action 
forces us to renormalize the theory in such a way that these 
gauge-variant local counterterms have a well-defined finite 
coefficient. An example of such a term is
\begin{equation}
\epsilon B_{VO} Tr(F_{LK} F_{SW}) Tr (F_{AG} F_{EN})
\epsilon^{VOLKSWAGEN}
\label{eq:VW}
\end{equation}
where $\epsilon = 1/720(2\pi)^5$. While this gives rise to 
many new interaction terms in the $d = 4$ theory, one possible
 manifestation seems to be of particular importance. 
Replacing one of the $TrF^2$ terms by their vev in 
extra dimensions, one arrives at
 \begin{equation}
\eta \epsilon^{\mu\nu\varrho\sigma} Tr (F_{\mu\nu} F_{\varrho\sigma})
\label{eq:FF}
\end{equation}
$\eta$ is the imaginary part of $T$, and unlike in the 
classical case it now (in addition to $\theta$)  couples to 
$F\tilde{F}$. Observe that (\ref{eq:FF}) is gauge-invariant, 
while (\ref{eq:VW}) is not, but is required by the absence of anomalies in 
$d = 10$. This shows that the remnants of such terms originate 
in ten dimensions, and are one of the few places where we 
could in principle observe whether we live in higher dimensions. 
(\ref{eq:FF}) suggests that not only $\theta$, but also $\eta$, couples 
like an axion. To make sure that this does not lead just to a 
redefinition of $\theta$ at the one-loop level, all anomaly 
cancellation terms have to be considered. Doing this and 
satisfying $TrF^2 = TrR^2$ in extra dimensions, one arrives 
at the result that $\eta$  couples differently to $E_6$ and $E_8^\prime$:
\begin{equation}
\epsilon\eta [(F\tilde{F})_8 - (F\tilde{F})_6]\\\\\
{\rm \\\ and\\\ }\\\\\
\theta [(F\tilde{F})_8 + (F\tilde{F})_6]
\end{equation}
leading to different gauge kinetic functions
\begin{equation}
f=S\pm\epsilon T
\label{eq:st}
\end{equation}
for the different gauge groups.

This fact has interesting consequences, some of which we will now list.
\begin{itemize}
\item[a)]
The second axion could be a candidate to solve the strong CP 
problem of QCD in the observable sector. One axion 
(like $\theta$ alone) would not be sufficient, because 
it is used to adjust the $\theta$-angle of $E_8^\prime$ and 
becomes massive. For a
relatively recent discussion see \cite{GKN}. 
\item[b)] Supersymmetry requires the same behaviour of the real
 parts of $S$ and $T$ as that of the imaginary part; i.e., 
$ReS$ and $ReT$ couple differently to $E_6$ and $E_8^\prime$. 
Since the vevs of these fields define the gauge coupling constants, 
$g_6$ and $g_8^\prime$ need no longer be equal. This might have 
consequences for the condensation scale of $E_8^\prime$.
\item[c)] There exist now two axion-dilaton pairs, and 
this might generalize the relaxation of the cosmological 
constant to the observable sector in the same way as is 
appears in the hidden sector \cite{Kim:2002tq}.
\item[d)] Imposition of supersymmetry also requires new terms 
in the K\"ahler potential at the one-loop level. We will 
discuss this later.
\item[e)] As expected, these effects at the one-loop level 
lead to an induced breakdown of supersymmetry in the 
observable sector once it is broken in the hidden sector. 
Remember our discussion in Section 5, where the observable 
sector remained supersymmetric. Gaugino masses are given by
\begin {equation}
m_{1/2} \sim F_T f_T + F_S f_S.
\end{equation}
At tree level we had $F_S = f_T = 0$ and vanishing gaugino masses. 
But now we have 
$f = S+\epsilon T$, and $f_T$ no longer  vanishes. As a 
result, non-trivial gaugino masses (and also non-trivial 
scalar masses and A-parameters) of order $\epsilon m_{3/2}$ 
are transmitted to the observable sector. 
\end{itemize}

Of course, in case of a nontrivial $\epsilon$, we should go back
to the action and see what happens to the one-loop effective 
potential. In general we shall expect a nontrivial value
of the cosmological constant (typically anti-de Sitter) and
in some cases even unbroken supersymmetry. 
So a disscussion of a fully realistic 
model needs more structure than present in the toy model under
consideration.


\section{Heterotic M-Theory}

With the discovery of string dualities,
there has been a revival of the study of those string theories that
might eventually become relevant for our discussion of
the low-energy effective supergravity theories.
From all the new and interesting results in string dualities,
it is the heterotic M--theory of Ho\v{r}ava and Witten
\cite{HW} (that in $d=11$ could be regarded as the
strong coupling limit of $d=10$ $E_8\times E_8$ heterotic string
theory) which
might have a direct impact on the discussion
of the phenomenological aspects of these theories.
One of the results concerns the question of the unification
of all fundamental coupling constants \cite{W} and the second
one the properties of the soft terms (especially the gaugino
masses) once supersymmetry is broken \cite{NOY,NOY2}. 
As we shall see in both cases, results
that appear problematic in the weakly
coupled case (as the formerly discussed heterotic string case
will be called from now on)
get modified in a satisfactory way, while the
overall qualitative picture remains essentially unchanged.
In these lectures we shall therefore concentrate on these aspects
of the new picture.

The heterotic M--theory is an 11--dimensional theory with the
$E_8\times E_8$ gauge fields living on two 10--dimensional
boundaries (walls), respectively, while the gravitational fields
can propagate in the bulk as well. A $d=4$ dimensional theory
with $N=1$ supersymmetry emerges at low energies when 6 dimensions 
are compactified on a Calabi--Yau manifold. The scales of that theory
are $M_{11}$, the $d=11$ Planck scale, $R_{11}$ the size of
the $x^{11}$ interval, and $V\sim R^6$ the volume of the
Calabi--Yau manifold. The quantities of interest in $d=4$,
the Planck mass, the GUT--scale and the unified gauge coupling
constant $\alpha_{GUT}$ should be determined through these
higher dimensional quantities. The fit of ref.\ \cite{W}
identifies $M_{GUT}\sim 3\cdot 10^{16}$ GeV  with the inverse 
Calabi--Yau radius $R^{-1}$. Adjusting $\alpha_{GUT}=1/25$ gives  
$M_{11}$ to be a few times larger than $M_{GUT}$. On the other hand, 
the fit of the actual value of the Planck scale can be achieved by
the choice of $R_{11}$ and, interestingly enough, $R_{11}$ turns out 
to be an order of magnitude larger than the fundamental 
length scale $M_{11}^{-1}$.
A satisfactory fit of the $d=4$ scales 
is thus possible, in contrast to the case of the weakly coupled
heterotic string where,
naively, the string scale seems to be a factor
20 larger than $M_{GUT}$.
%

%
\subsection{The action in $d=11$}

The effective action of the strongly coupled
$E_8 \times E_8$ -- $M$--theory in the ``downstairs'' 
approach is given by
\cite{HW} 
(we take into account the numerical corrections found in 
\cite{CC})
\bea
L
\!\!&=&\!\!
{1\over \kappa^2} \int
d^{11}x \sqrt{g}
\left[
       - \frac{1}{2}R
       - \frac{1}{2} \psib_I \Gamma^{IJK} D_J \psi_K
       - \frac{1}{48} G_{IJKL} G^{IJKL}
\right.
\nn\\
&&
       - \frac{\sqrt{2}}{384}
            \left( \psib_I \Gamma^{IJKLMN} \psi_N
                  +12 \psib^J \Gamma^{KL} \psi^M \right)
            \left( G_{JKLM} + {\hat G}_{JKLM} \right)
\nn\\
&&\qquad\quad
       - \left. \frac{\sqrt{2}}{3456}
            \epsilon^{I_1 I_2 \ldots I_{11}} C_{I_1 I_2 I_3}
            G_{I_4 \ldots I_7} G_{I_8 \ldots I_{11}}
\right]
\\
\!\!&+&\!\!
\frac{1}{4\pi(4\pi\kappa^2)^{2/3}}  \int_{M^{10}_i} 
d^{10}x \sqrt{g}
\left[ 
    - \frac{1}{4} F^a_{iAB} F_i^{aAB}
       - \frac{1}{2} \chib_i^a \Gamma^AD_A ({\hat\Omega}) \chi_i^a
\right.
\nn\\
&&
\left.
       - \frac{1}{8} \psib_A \Gamma^{BC} \Gamma^A
            \left( F^a_{iBC} + {\hat F}^a_{iBC} \right) \chi_i^a
       + \frac{\sqrt{2}}{48}
            \left( \chib_i^a \Gamma^{ABC} \chi_i^a\right){\hat G}_{ABC11}
\right]
\nn
\eea
where $M^{11}$ is the $d=11$ manifold 
and $M_i^{10}$  its 10--dimensional boundaries. 
In the lowest approximation $M^{11}$ is just
a product $M^4 \times X^6 \times S^1/Z_2$.
Compactifying to $d=4$ in such an approximation we obtain
\cite{W,CC}
\be
G_N 
= {\kappa_4^2\over 8 \pi} 
= {\kappa^2 \over 8 \pi R_{11} V}
\,,
\label{eq:GN}
\ee
\be
\alpha_{GUT} = {(4\pi\kappa^2)^{2/3} \over V}
\label{eq:alphaGUT}
\ee
with $V$ the volume of the Calabi--Yau manifold $X^6$
and $R_{11} = \pi\rho$ the $S^1/Z_2$ length.

The fundamental mass scale of the 11--dimensional theory is given by 
$M_{11} = \kappa^{-2/9}$. Let us see which value of $M_{11}$ is
favoured in a phenomenological application. For that purpose we
identify the Calabi--Yau volume $V$ with the GUT--scale: 
$V\sim(M_{GUT})^{-6}$. From (\ref{eq:alphaGUT}) and the value of
$\alpha_{GUT}=1/25$ at the grand unified scale, we can then deduce
the value of $M_{11}$
\be
V^{1/6} M_{11} 
=
(4\pi)^{1/9} \alpha_{GUT}^{-1/6}
\approx
2.3
\,,
\label{eq:VM11}
\ee
to be a few times larger than the GUT--scale. In a next step we
can now adjust the gravitational coupling constant by choosing
the appropriate value of $R_{11}$ using (\ref{eq:GN}). 
This leads to
\be
R_{11} M_{11}
=
\left(\frac{M_{Planck}}{M_{11}}\right)^2
\frac{\alpha_{GUT}}{8\pi(4\pi)^{2/3}}
\approx
2.9 \cdot 10^{-4} \left(\frac{M_{Planck}}{M_{11}}\right)^2
\,.
\label{eq:R11M11}
\ee
This simple analysis tells us the following: 

\begin{itemize}

\item
In contrast to the
weakly coupled case,
the correct value of $M_{\rm Planck}$ can be fitted by adjusting
the value of $R_{11}$.

\item
The numerical value of $R_{11}^{-1}$ turns out to be 
approximately an order of magnitude smaller than $M_{11}$.

\item
Thus the 11th dimension appears to be larger than the dimensions
compactified on the Calabi--Yau manifold, and at an intermediate
stage the world appears 5--dimensional with two 4--dimensional
boundaries (walls).

\end{itemize}

We thus have the following picture of the evolution and unification
of coupling constants. 
At low energies the
world is 4--dimensional and the couplings evolve accordingly with
energy: a logarithmic variation of gauge coupling constants and
the usual power law behaviour for the gravitational coupling.
Around $R_{11}^{-1}$ we have an additional 5th dimension and the
power law evolution of the gravitational interactions changes.
Gauge couplings are not effected at that scale since the
gauge fields live on the walls and do not feel the existence of
the 5th dimension. Finally at $M_{GUT}$ the theory becomes
11--dimensional and both gravitational and gauge couplings
show a power law behaviour and meet at the scale $M_{11}$,
the fundamental scale of the theory. It is obvious that  
the correct choice of $R_{11}$ is needed to achieve unification.
We also see that, although the theory is weakly coupled at
$M_{GUT}$, this is no longer true at $M_{11}$. The naive
estimate for the evolution of the gauge coupling constants between
$M_{GUT}$ and $M_{11}$ goes with the sixth power of the scale.
At $M_{11}$ we thus expect unification of the couplings
at $\alpha\sim O(1)$. In that sense, the M--theoretic description
of the heterotic string gives an interpolation between
weak coupling and moderate coupling. In $d=4$ this is not
strong--weak coupling duality in the usual sense. We shall later
come back to these questions when we discuss the appearance of
a critical
limit on the size of $R_{11}$. A value of $\alpha\sim O(1)$
(and thus $S\sim O(1)$) at $M_{11}$ might also be favoured in
view of the question of the dynamical determination of the vev
of the dilaton field \cite{3}.
%
%

\subsection{The effective action in $d=4$}

 We now want to perform a compactification from $d=11$ 
to $d=4$. Again we  use the method of reduction and truncation. 
For the metric we write
\be
g^{(11)}_{MN} =
\left(
\ba{ccc}
c_4 e^{-\gamma} e^{-2\sigma} g_{\mu\nu} & & \\
 & e^\sigma g_{mn} & \\
 & & e^{2\gamma} e^{-2\sigma}
\ea
\right)
\label{eq:g11}
\ee
with 
$M,N = 1 \ldots 11$; $\mu,\nu = 1 \ldots 4$; $m,n = 5 \ldots 10$ 
and det($g_{mn}$)=1. 
This is the frame in which the 11--dimensional Einstein action 
gives the ordinary Einstein action after the reduction do $d=4$:
\be
-\frac{1}{2\kappa^2} \int d^{11}x \sqrt{g^{(11)}} R^{(11)} 
=
-\frac{c_4 \hat{V_7}}{2\kappa^2} \int d^{4}x \sqrt{g} R + \ldots
\ee
where $\hat{V_7}=\int d^7x$ is the coordinate volume of the 
compact 7--manifold and the scaling factor $c_4$ describes our 
freedom to choose the units in $d=4$. The most popular choice in the 
literature is $c_4=1$. This, however, corresponds to the unphysical 
situation in which the 4--dimensional Planck mass is determined 
by the choice of $\hat{V_7}$ which is just a convention. With 
$c_4=1$ one needs further rescaling of the 4--dimensional metric. 
We instead prefer the choice
\be
c_4 = V_7 / \hat{V_7}
\ee
where $V_7 = \int d^7x \sqrt{g^{(7)}}$ is the physical volume 
of the compact 7--manifold. This way we recover eq.\ (\ref{eq:GN}) 
in which the 4--dimensional Planck mass depends on the physical 
(and not coordinate) volume of the manifold on which we compactify. 
As a result,
if we start from the product of the 4--dimensional Minkowski 
space and some 7--dimensional compact space  
(in the leading order of the expansion in $\kappa^{2/3}$) 
as a ground state in $d=11$ 
we obtain the Minkowski space with the standard normalization 
as the vacuum in $d=4$.

To find a more explicit formula for $c_4$ we have to discuss 
the fields $\sigma$ and $\gamma$ in some detail. In the 
leading approximation $\sigma$ is the 
overall modulus of the Calabi--Yau 6--manifold. We can divide 
it into a sum of the vacuum expectation value, 
$\left< \sigma \right>$, and the fluctuation $\tilde\sigma$.
In general both parts could depend on all 11 coordinates 
but in practice we have to impose some restrictions. 
The vacuum expectation value can not depend on $x^\mu$ 
if the 4--dimensional theory is to be Lorentz--invariant. 
In the fluctuations we drop the dependence on the compact 
coordinates corresponding to the higher Kaluza--Klein modes. 
Furthermore, we know that in the leading approximation 
$\left< \sigma \right>$ is just a constant, $\sigma_0$ , 
while corrections depending on the internal coordinates,
$\sigma_1$, are of the next order in $\kappa^{2/3}$. 
Thus, we obtain
\be
\sigma(x^\mu,x^m,x^{11})
=
\left<\sigma\right>(x^m,x^{11})
+ \tilde{\sigma}(x^\mu)
=
\sigma_0 
+ \sigma_1(x^m,x^{11}) 
+ \tilde{\sigma}(x^\mu)
\,.
\label{eq:sigma}
\ee
To make the above decomposition unique we define $\sigma_0$ 
by requiring that the integral of $\sigma_1$ over the internal 
space vanishes. The analogous decomposition can be also done 
for $\gamma$. With the above definitions the physical volume 
of the compact space is
\be
V_7 
= \int d^7x \left<e^{2\sigma}e^{\gamma}\right>
= e^{2\sigma_0} e^{\gamma_0} \hat{V_7}
\ee
up to corrections of order $\kappa^{4/3}$. Thus, the parameter  
$c_4$ can be written as
\be
c_4 = e^{2\sigma_0}e^{\gamma_0}
\,.
\ee

The choice of the coordinate volumes is just a convention. 
For example in the case of the Calabi--Yau 6--manifold only 
the product $e^{3\sigma}\hat{V_6}$ has physical meaning. 
For definiteness we will use the convention that the coordinate 
volumes are equal 1 in $M_{11}$ units. Thus, 
$\left<e^{3\sigma}\right>$ describes 
the Calabi--Yau volume in these units. 
Using eqs.\ (\ref{eq:VM11},\ref{eq:R11M11}) 
we obtain $e^{3\sigma_0} = V M_{11}^6 \approx (2.3)^6$, 
$e^{\gamma_0} e^{-\sigma_0} = R_{11} M_{11} \approx 9.2 a^2$. 
The parameter $c_4$ is equal to the square of the 4--dimensional 
Planck mass in these units and numerically $c_4 \approx (35a)^2$.

At the classical level we compactify on 
$M^4 \times X^6 \times S^1/Z_2$. 
This means that the vacuum expectation values 
$\left<\sigma\right>$ and $\left<\gamma\right>$ 
are just constants and eq.\ (\ref{eq:sigma}) reduces to
\be
\sigma = \sigma_0 + \tilde\sigma(x^\mu)
,\qquad\qquad
\gamma = \gamma_0 + \tilde\gamma(x^\mu)
\,.
\ee
In such a situation $\sigma$ and $\gamma$ are 4--dimensional fields.
We introduce two other 4--dimensional fields by the relations
\bea
\frac{1}{4! c_4} e^{6\sigma} G_{11\lambda\mu\nu} &=& 
\epsilon_{\lambda\mu\nu\rho}\left(\partial^\rho D \right)
\,,
\\
C_{11 a {\bar b}} &=& C_{11} \delta_{a \bar b}
\eea
where $x^a$ ($x^{\bar b}$) is the holomorphic (antiholomorphic) coordinate 
of the Calabi--Yau manifold. 
Now we can define the dilaton and the modulus fields by
\bea
\sS &=& \frac{1}{\left( 4\pi \right)^{2/3}} 
\left( e^{3\sigma} + i 24 \sqrt{2} D \right)
\,,
\label{eq:sS0}
\\
\sT &=& \frac{1}{\left( 4\pi \right)^{2/3}}
\left( e^{\gamma} + i 6 \sqrt{2} C_{11} + C^*_i C_i \right)
\label{eq:sT0}
\eea
where the observable sector matter fields $C_i$ originate from the
gauge fields $A_M$ on the 10--dimensional observable wall 
(and $M$ is an index in
the compactified six dimensions).
The K\"ahler potential takes its standard form
\be
K = - \log (\sS + \sS^*) - 3 \log (\sT+\sT^* -2 C^*_i C_i)
\label{eq:sK}
\,.
\ee
The imaginary part of $\sS$ (Im$\sS$) corresponds to the model 
independent axion, and with the above normalization the gauge kinetic 
function is $f = \sS$. We have also
\be
W(C) = d_{ijk} C_iC_jC_k   
\ee
Thus the action to leading order is very similar to the weakly coupled case.

Before drawing any conclusion from the formulae obtained above we 
have to discuss a possible obstruction at the next to leading order. 
For the 3--index tensor field $H$ in $d=10$ supergravity to be well 
defined one has to satisfy $dH = \tr F_1^2 + \tr F_2^2 - \tr R^2 = 0$ 
cohomologically. In the simplest case of the standard embedding one
assumes $\tr F_1^2 = \tr R^2$ locally and the gauge group is broken to 
$E_6 \times E_8$. Since in the M--theory case the two different gauge 
groups live on the two different boundaries (walls) of space--time 
such a cancellation point by point is no longer possible 
\cite{W}.
We expect nontrivial vacuum expectation values (vevs) of
\be
(dG) \propto \sum_i \delta(x^{11} - x^{11}_i) 
\left( \tr F_i^2 - {1\over 2} \tr R^2 \right)
\ee
at least on one boundary ($x^{11}_i$ is the position of $i$--th 
boundary). In the case of the standard embedding we would have 
$\tr F_1^2 - {1\over 2} \tr R^2 = {1\over 2} \tr R^2$ on one and 
$\tr F^2_2 - {1\over 2} \tr R^2 = - {1\over 2} \tr R^2$ on the other
boundary. This might pose a severe problem since a nontrivial vev  of 
$G$ might be in conflict with supersymmetry ($G_{11ABC}=H_{ABC}$). 
The supersymmetry transformation law in $d=11$ reads  
\be
\delta \psi_M
=
D_M\eta + \frac{\sqrt{2}}{288} G_{IJKL} 
          \left( \Gamma_M^{IJKL} - 8 \delta_M^I \Gamma^{JKL} 
            \right) \eta
+ \ldots
\label{eq:dpsiM}
\ee
Supersymmetry will be broken unless e.g.\ the derivative term
$D_M\eta$ compensates the nontrivial vev of $G$. Witten has shown 
\cite{W} 
that such a cancellation can occur and constructed the solution in 
the linearized approximation 
(linear in the expansion parameter $\kappa^{2/3}$).
This solution requires some modification of the metric on $M^{11}$:
\be
g^{(11)}_{MN} =
\left(
\ba{ccc}
(1+b) \eta_{\mu\nu} & & \\
 & (g_{ij}+h_{ij}) & \\
 & & (1+\gamma') 
\ea
\right)
\,.
\label{eq:gW}
\ee
$M^{11}$ is no longer a direct product $M^4 \times X^6 \times S^1/Z_2$ 
because $b$, $h_{ij}$ and $\gamma'$ depend
now on the compactified coordinates.
The volume of $X^6$ depends on $x^{11}$ 
\cite{W}:
\be
\frac{\partial}{\partial x^{11}} V 
=
-\frac{\sqrt{2}}{8}
{\int d^6x \sqrt{g} \omega^{AB}\omega^{CD}G_{ABCD}}
\label{eq:d11V_W}
\ee
where the integral is over the Calabi--Yau manifold $X^6$ and $\omega$
is the corresponding K\"ahler form.
The parameter $(1+b)$ is the scale factor of
the Minkowski 4--manifold and depends on $x^{11}$ in the following way
\be
\frac{\partial}{\partial x^{11}} b =
\frac{1}{2}  \frac{\partial}{\partial x^{11}} \log v_4 =
\frac{\sqrt{2}}{24}
\omega^{AB}\omega^{CD}G_{ABCD}
\label{eq:d11b}
\ee
where $v_4$ is the physical volume for some fixed coordinate volume
in $M^4$.
In our simple reduction and truncation method
with the metric $g^{(11)}_{MN}$ given by eq.\ (\ref{eq:g11}) 
we can reproduce the $x^{11}$ dependence of $V$ and $v_4$.
The volume of $X^6$ is determined by $\sigma$:
\be
\frac{\partial}{\partial x^{11}} \log V 
= \frac{\partial}{\partial x^{11}} \left(3 \left<\sigma\right>\right)
= 3 \frac{\partial}{\partial x^{11}} \sigma
\label{eq:d11V}
\ee
while the scale factor of $M^4$ can be similarly
expressed in terms of $\sigma$ and $\gamma$ fields:
\be
\frac{\partial}{\partial x^{11}} \log v_4 
= - \frac{\partial}{\partial x^{11}} 
\left(2 \left<\gamma\right> + 4 \left<\sigma\right>\right)
= - \frac{\partial}{\partial x^{11}} (2 \gamma + 4 \sigma)
\label{eq:d11v4}
\,.
\ee
Substituting $\left<\sigma\right>$ with $\sigma$ in the above 
two equations is allowed because, due to our decomposition 
(\ref{eq:sigma}), only the vev of $\sigma$ depends on the 
internal coordinates (the same is true for $\gamma$).
The scale factor $b$ calculated in ref.\ \cite{W}
depends also on the Calabi--Yau coordinates.
Such a dependence can not be reproduced in our simple
reduction and truncation compactification so we have to average
eq.\ (\ref{eq:d11b}) over $X^6$.
Using equations (\ref{eq:d11V_W}--\ref{eq:d11v4})
after such an averaging we obtain
(to leading order in the expansion parameter $\kappa^{2/3}$) 
\cite{NOY}
\be
\frac{\partial\gamma}{\partial x^{11}} 
=
-\frac{\partial\sigma}{\partial x^{11}} 
=
\frac{\sqrt{2}}{24}
\frac
{\int d^6x \sqrt{g} \omega^{AB}\omega^{CD}G_{ABCD}}
{\int d^6x \sqrt{g}}
\,.
\label{eq:sigmagamma1}
\ee
Substituting the vacuum expectation value of $G$ found in 
\cite{W}
we can rewrite it in the form
\be
\frac{\partial\gamma}{\partial x^{11}} 
=
-\frac{\partial\sigma}{\partial x^{11}} 
=
\frac{2}{3} \alpha \kappa^{2/3} V^{-2/3}
\label{eq:sigmagamma2}
\ee
where 
\be
\alpha = \frac{\pi c}{2(4\pi)^{2/3}}
\label{eq:alpha}
\ee
and c is a constant of order unity given for the standard embedding 
of the spin connection by
\be
c =  V^{-1/3}
\left| \int \frac{\omega \wedge \tr (R \wedge R)}{8 \pi^2} \right|
\,.
\label{eq:c}
\ee
Our calculations, as those of Witten, are valid only in the 
leading nontrivial order in the $\kappa^{2/3}$ expansion. 
The expression (\ref{eq:sigmagamma2}) for the derivatives 
of $\sigma$ and $\gamma$ contain an explicit factor $\kappa^{2/3}$. 
This means that we should take the lowest order value for the 
Calabi--Yau volume in that expression. An analogous procedure 
has been used in obtaining all formulae presented in this paper. 
We always expand in $\kappa^{2/3}$ and drop all terms which are 
of higher order. Taking the above into 
account and using our units in which $M_{11}=1$ we can rewrite 
eq.\ (\ref{eq:sigmagamma2}) in the simple form:
\be
\frac{\partial\gamma}{\partial x^{11}} 
=
-\frac{\partial\sigma}{\partial x^{11}} 
=
\frac{2}{3} \alpha e^{-2\sigma_0}
\,.
\label{eq:sigmagamma3}
\ee

Eqs.\ (\ref{eq:sigmagamma1}--\ref{eq:sigmagamma3}) as
derived in ref.\ \cite{NOY}
 contain all the information to deduce
the effective action, i.e. K\"ahler potential,
superpotential and gauge kinetic function of the 4--dimensional 
effective supergravity theory. 

It is the above dependence of $\sigma$ and $\gamma$ on $x^{11}$
that leads to these  consequences. 
One has to be careful in defining the fields in $d=4$. It is obvious,
that the 4--dimensional fields $\sS$ and $\sT$ can not be any 
longer defined by eqs.\ (\ref{eq:sS0}, \ref{eq:sT0}) 
because now $\sigma$ and $\gamma$ are 5--dimensional fields. 
We have to integrate out the dependence on the 11th coordinate. 
In the present approximation, this procedure is quite simple:
we have to replace $\sigma$ and $\gamma$ in the 
definitions of $\sS$ and $\sT$ with their averages over 
the $S^1/Z_2$ interval \cite{NOY}. 
With the linear dependence of $\sigma$ and $\gamma$ on $x^{11}$ 
their average values coincide with the values taken at the 
middle of the $S^1/Z_2$ interval
\be
\bar \sigma 
= \sigma \left( \frac{\pi\rho}{2} \right)
= \sigma_0 + \tilde{\sigma}(x^\mu)
\,,
\ee
\be
\bar \gamma 
= \gamma \left( \frac{\pi\rho}{2} \right)
= \gamma_0 + \tilde{\gamma}(x^\mu)
\,.
\ee

When we reduce the boundary part of the Lagrangian of M--theory 
to 4 dimensions we find exponents of $\sigma$ and $\gamma$ 
fields evaluated at the boundaries. Using eqs.\ (\ref{eq:sigma}) 
and (\ref{eq:sigmagamma3}) we get
\bea
e^{-\gamma} \big|_{M^{10}_i}
&=&
e^{-\gamma_0} \pm \frac{1}{3} \alpha e^{-3\sigma_0}
\,,
\label{eq:exp-gamma}
\\
e^{3\sigma} \big|_{M^{10}_i}
&=&
e^{3\sigma_0} \pm \alpha e^{\gamma_0}
\,.
\label{eq:exp3sigma}
\eea
The above formulae have very important consequences for 
the definitions of the K\"ahler potential and the 
gauge kinetic functions. For example, the coefficient in front of the 
$D_\mu C^*_i D^\mu C_i$ kinetic term is proportional to $e^{-\gamma}$ 
evaluated at the $E_6$ wall where the matter fields propagate. 
At the lowest order this was just $e^{-\gamma_0}$ or 
$\left<\sT\right>^{-1}$ up to some numerical 
factor. From eq.\ (\ref{eq:exp-gamma}) we see that 
at the next to leading order also $\left<\sS\right>^{-1}$ 
is involved with relative coefficient $\alpha/3$. 
Taking such corrections into account we find that 
at this order the K\"ahler potential is given 
by
\be
K 
= 
- \log (\sS + \sS^*) 
+ \frac{2 \alpha C^*_i C_i}{\sS + \sS^*}
- 3 \log (\sT+ \sT^* - 2 C^*_i C_i)
\ee
with $\sS$ and $\sT$ now defined by
\bea
\sS &=& \frac{1}{\left( 4\pi \right)^{2/3}}
\left( e^{3\bar\sigma} + i 24 \sqrt{2} \bar D + \alpha C^*_i C_i \right)
\,,
\\
\sT &=& \frac{1}{\left( 4\pi \right)^{2/3}}
\left( e^{\bar\gamma} + i 6 \sqrt{2} \bar C_{11} + C^*_i C_i \right)
\label{eq:sT}
\eea
where bars denote averaging over the 11th dimension. 
It might be of some interest to note that the combination
$\sS\sT^3$ is independent of $x^{11}$ 
even before this averaging procedure took place.
The solution above is valid only for terms at most linear in $\alpha$. 
Keeping this in mind we could write the K\"ahler potential 
also in the form 
\be
K= 
-\log(\sS + \sS^*- 2 \alpha C^*_i C_i)
- 3 \log (\sT+ \sT^* - 2 C^*_i C_i).
\ee

Equipped with this definition the calculation of the gauge kinetic
function(s) from eqs.\ (\ref{eq:sigmagamma3}, \ref{eq:exp3sigma})
becomes a trivial exercise \cite{NOY}. In the five--dimensional
theory $f$ depends on the 11--dimensional coordinate as well, thus the 
gauge kinetic function takes different values at the two walls. 
The averaging procedure allows us to deduce these functions directly. 
For the simple case at hand (the so--called standard embedding)
eq.\ (\ref{eq:exp3sigma}) gives \cite{NOY}
\be
f_6 = \sS +\alpha \sT\,;
\qquad\qquad
f_8 = \sS - \alpha \sT\,.
\label{eq:f6f8alpha}
\ee
It is a special property of the standard embedding that the 
coefficients are equal and opposite. The coefficients might vary for 
more general cases. This completes the discussion of the $d=4$ 
effective action in next to leading order, noting that the 
superpotential does not receive corrections at this level.

The nontrivial dependence of $\sigma$ and $\gamma$ on $x^{11}$
can also enter definitions and/or interactions of other
4--dimensional fields. Let us next consider the gravitino. 
After all we have to show that this field is massless to give 
the final proof that the given solution respects supersymmetry.
Its 11--dimensional kinetic term
\be
- \frac{1}{2} \sqrt{g} \psib_I \Gamma^{IJK} D_J \psi_K
\label{eq:grav_kin}
\ee
remains diagonal after compactification to $d=4$ if we
define the 4--dimensional gravitino, $\psi^{(4)}_\mu$,
and dilatino, $\psi^{(4)}_{11}$, fields by the relations
\bea
\psi_\mu
&=&
e^{-(\sigma-\sigma_0)/2}e^{-(\gamma-\gamma_0)/4}
\left(\psi^{(4)}_\mu + \frac{1}{\sqrt{6}}\Gamma_\mu\psi^{(4)}_{11}\right)
\,,
\label{eq:grav_def1}
\\
\psi_{11}
&=&
-\frac{2}{\sqrt{6}}e^{(\sigma-\sigma_0)/2} e^{(\gamma-\gamma_0)/4}
\Gamma^{11} \psi^{(4)}_{11}
\,.
\label{eq:grav_def2}
\eea
The $d=11$ kinetic term (\ref{eq:grav_kin}) gives after
the compactification also a mass term for the $d=4$ gravitino
of the form
\be
\frac{3}{8} e^{\sigma_0} e^{-\gamma_0}
\frac{\partial\gamma}{\partial x^{11}}
=
\frac{\sqrt{2}}{64} e^{\sigma_0} e^{-\gamma_0}
\frac
{\int d^6x \sqrt{g} \omega^{AB}\omega^{CD}G_{ABCD}}
{\int d^6x \sqrt{g}}
=
\frac{1}{4} \alpha e^{-\sigma_0} e^{-\gamma_0}
\,.
\label{eq:grav_mass}
\ee
The sources of such a term are nonzero values of the spin 
connection components $\omega_\mu^{\alpha 11}$ and 
$\omega_m^{a 11}$ resulting from the $x^{11}$ dependence of 
the metric. It is a constant mass term from the 4--dimensional 
point of view. This, however, does not mean that the gravitino 
mass is nonzero. There is another contribution from the 
11--dimensional term
\be
- \frac{\sqrt{2}}{384} \sqrt{g}
\psib_I \Gamma^{IJKLMN} \psi_N
\left( G_{JKLM} + {\hat G}_{JKLM} \right)
\,.
\label{eq:GG}
\ee
After redefining fields according to
(\ref{eq:grav_def1},\ref{eq:grav_def2})
and averaging the nontrivial vacuum expectation value of $G$
over $X^6$ we get from eq.\ (\ref{eq:GG})
a mass term which exactly cancels the previous contribution
(\ref{eq:grav_mass}).
The gravitino is massless -- the result which we expect
in a model with unbroken supersymmetry and vanishing
cosmological constant.
Thus, we find that our simple reduction and truncation method
(including the correct $x^{11}$ dependence in next to leading order)
reproduces the main features of the model.

The factor $\left<\exp(3\sigma)\right>$ represents the  volume of the
six--dimensional compact space in units of $M_{11}^{-6}$.
The $x^{11}$ dependence of $\sigma$
then leads to the geometrical picture that the volume of this space
varies with $x^{11}$ and differs at the two boundaries:
\be
V_{E_8}
=
V_{E_6} - 2 \pi^2 \rho \left(\frac{\kappa}{4 \pi}\right)^{2/3}
\left|
\int \omega \wedge
     \frac{\tr (F \wedge F) - \frac{1}{2} \tr (R \wedge R)}{8 \pi^2}
\right|
\ee
where the integral is over $X^6$ at the $E_6$ boundary.
In the given approximation, this variation is linear,
and for growing $\rho$ the volume on the $E_8$ side becomes
smaller and smaller. At a critical value of $\rho$
the volume will thus vanish and this will provide
us with an upper limit on $\rho$:
\be
\rho
<
\rho_{crit}
=
\frac{(4 \pi)^{2/3}}{c\pi^2} M_{11}^{3} V_{E_6}^{2/3}
\label{eq:rho}
\ee
where $c$ was defined in eq.\ (\ref{eq:c}). The 
critical value is model dependent and we shall not discuss this
in detail here. 

Let us now compare the M-theory picture with that of the weakly
coupled heterotic string.
Inspection of (\ref{eq:st}) and (\ref{eq:f6f8alpha}) reveals a close
connection between the two \cite{BD,NS}.
The variation of the Calabi--Yau manifold volume as discussed above
is the analogue of the one loop correction of the gauge kinetic 
function (\ref{eq:st}) 
in the weakly coupled case and has the same 
origin, namely a Green--Schwarz anomaly cancellation counterterm. 
In fact, also in the strongly coupled case this leads to a correction
for the gauge coupling constants at the $E_6$ and $E_8$ side. 
As seen, gauge couplings are no longer given by the (averaged) 
$\sS$--field, but by that combination of the (averaged) $\sS$ and 
$\sT$ fields which corresponds to the $\sS$--field before averaging 
at the given boundary leading to 
\be
f_{6,8} = \sS \pm \alpha \sT
\ee
at the $E_6$ ($E_8$) side respectively.
The critical value of $R_{11}$ will correspond to infinitely strong
coupling at the $E_8$ side $\sS - \alpha \sT = 0$.
Since we are here close to criticality a correct phenomenological  
fit of $\alpha_{\rm GUT} = 1/25$ should include this correction
$\alpha_{\rm GUT}^{-1} = \sS + \alpha \sT$ where $\sS$ and
$\alpha \sT$ give comparable contributions. This is a difference to 
the weakly coupled case, where in $f= \wS + \epsilon \wT$ the latter
contribution was small compared to $\wS$. The stability
of this  result for the 
corrections to $f$ when going from weak coupling to strong coupling 
is only possible because of the rather special properties of $f$. 
$f$ does not receive further perturbative corrections beyond one loop
\cite{SV,N}, and the one loop corrections are determined by
the anomaly considerations. The formal expressions for the
corrections are identical, the difference being only that in the
strongly coupled case these corrections are 
to be interpreted of comparable importance as
the classical value.

%
%
\section{Supersymmetry breaking at the hidden wall}

For the discussion of supersymmetry breakdown we should
carefully examine the supersymmetry transformation of fermionic
fields. Of particular importance are the fields that originate from
the higher dimensional gravitino.
For the $d=11$ action, the
supersymmetry transformation laws for 
these  fields  are given by
\begin{eqnarray}
  \delta \psi_A &=& D_A\eta + 
      \frac{\sqrt{2}}{288} G_{IJKL} \left(
    \Gamma_A^{IJKL} - 8 \delta_A^I \Gamma^{JKL} \right) \eta - 
\label{eq:susytr-strong-A}
  \\ 
   &-&  \frac{1}{1152\pi} \left( \frac{\kappa}{4\pi} \right)^{2/3} 
  \delta(x^{11}) \left( \bar \chi^a \Gamma_{BCD} \chi^a \right) \left
    ( \Gamma_A^{BCD} - 6 \delta_A^B \Gamma^{CD} \right) \eta \ldots 
  \nonumber 
\end{eqnarray}

as well as
\begin{eqnarray}
  \delta \psi_{11} &=& D_{11} \eta + \frac{\sqrt{2}}{288} G_{IJKL}
  \left( \Gamma_{11}^{IJKL} - 8 \delta_{11}^I \Gamma^{JKL} \right)
  \eta + \nonumber \\ &+& \frac{1}{1152\pi} \left( \frac{\kappa}{4\pi}
  \right)^{2/3} \delta(x^{11}) \left( \bar \chi^a \Gamma_{ABC} \chi^a
  \right) \Gamma^{ABC} \eta + \ldots \label{eq:susytr-strong-B}
\end{eqnarray}
where gaugino bilinears appear in the right hand side of both
expressions. 
Again we consider gaugino condensation  at the hidden $E_8$ boundary 
\begin{equation}
\langle \bar{\chi}^a \Gamma_{ijk} \chi^a \rangle = g_8^2 \Lambda^3
\epsilon_{ijk}.
\end{equation}
The $E_8$ gauge coupling constant appears in this equation because 
the straightforward reduction and truncation leaves a non--canonical
normalization for the gaugino kinetic term. 
An important property of the  weakly coupled case (d=10
Lagrangian) was the fact that  the gaugino condensate and 
the three--index tensor field
$H$ contributed to the scalar potential in a full square.  
Ho\v{r}ava made the important observation that a
similar structure appears in the M--theory Lagrangian as well \cite{H}:
\begin{equation}
- \frac{1}{12\kappa^2} \int_{M^{11}} d^{11}x \sqrt{g}
  \left(G_{ABC11} 
     - \frac{\sqrt{2}}{32\pi} \left( \frac{\kappa}{4\pi} \right)^{2/3}
               \delta(x^{11}) \bar{\chi}^a \Gamma_{ABC} \chi^a
  \right)^2  \label{eq:perfect-square-strong}
\end{equation}
with the obvious relation between $H$ and $G$. Let us now have a closer
look at the form of $G$. At the next to leading order we have
\begin{eqnarray}
     G_{11ABC}&=&(\partial_{11} C_{ABC} +\mbox{permutations}) 
\nonumber \\
             &+&\frac{1}{4 \pi \sqrt{2}} 
              \left( \frac{\kappa}{4 \pi} \right)^{2/3}
             \sum_{i} \delta(x^{11}-x_i^{11})
          ( \omega^{YM}_{ABC}-\frac{1}{2} \omega^L_{ABC} ).
\end{eqnarray}
Observe, that in the bulk we have $G=dC$ with the Chern--Simons
contributions confined to the boundaries.  Formula 
(\ref{eq:perfect-square-strong}) suggests a cancellation between 
the gaugino condensate and the $G$--field in a way very similar to 
the weakly coupled case, but the nature of the cancellation of the 
terms becomes much more transparent now. 
Remember that
in the former case we had 
argued that because of 
the quantization condition for $<dB>$ the gaugino 
condensate is cancelled not by $<dB>$ but 
by a flux of the Chern--Simons terms. 
{\it Here this 
becomes obvious.} The condensate is located at the wall as are the 
Chern--Simons terms, so this cancellation has to happen 
locally at the wall and $dC$ should vanish for $G$ not to have
a vev in the bulk. In any case there is a quantization condition for
$dC$ as well \cite{WQ}.

So this cancellation is very similar to the one in the weakly 
coupled case.
At the minimum of the potential we obtain $G_{ABCD}=0$ everywhere and
\begin{equation}
G_{ABC11} 
     = \frac{\sqrt{2}}{32\pi} \left( \frac{\kappa}{4\pi} \right)^{2/3}
               \delta(x^{11}) \bar{\chi}^a \Gamma_{ABC} \chi^a
\label{eq:vev-G}
\end{equation}
at the hidden wall.
Eqs.~(\ref{eq:susytr-strong-A}) and (\ref{eq:susytr-strong-B}) 
then become
\begin{eqnarray}
    \delta \psi_A& =& D_A \eta +\ldots 
\\
  \delta \psi_{11} &=&  D_{11}\eta
               + \frac{1}{384\pi} \left( \frac{\kappa}{4\pi}
  \right)^{2/3} \delta(x^{11}) \left( \bar \chi^a \Gamma_{ABC} \chi^a
  \right) \Gamma^{ABC} \eta \ldots \label{eq:susy-tr}
\end{eqnarray}
An inspection of the potential shows that $\delta \psi_{11}$ is
nonvanishing and supersymmetry is spontaneously broken.
Because of
the cancellation in eq.~(\ref{eq:perfect-square-strong}), the
cosmological constant vanishes
to leading order.  Recalling supersymmetry transformation
law for the elfbein
\begin{equation}
    \delta e^m_I=\frac{1}{2}\bar \eta \Gamma^m \psi_I,
\end{equation}
one finds that the superpartner of the $\sT$ field plays the role 
of the goldstino. Again we have a situation where $F_{\sS}=0$ 
(due to the cancellation in (\ref{eq:perfect-square-strong})) with
nonvanishing $F_{\sT}$. But here we find the novel and interesting
situation that $F_{\sT}$ differs from zero only at the hidden wall,
although the field itself is a bulk field. 

At that wall our discussion is  completely 4--dimensional although
we are still dealing effectively with a $d=5$ theory. To reach
the effective theory in $d=4$ we have to integrate out the 
dependence of the $x^{11}$ coordinate. As in the previous section
this can be performed by the averaging procedure explained there.
With the gaugino condensation scale $\Lambda$ sufficiently small 
compared to the
compactification scale $M_{GUT}$, the low--energy effective theory is
well described by four dimensional $N=1$ supergravity in which
supersymmetry is spontaneously broken.  In this case, the modes which
remain at low energies will be well approximated by constant modes
along the $x^{11}$ direction.  This observation justifies our
averaging procedure to obtain four dimensional quantities. 
Averaging $\delta \psi_{11}$ over $x^{11}$, we thus obtain the vev
of the auxiliary field $F_{\sT}$
\begin{equation}
    F_{\sT}=\frac{1}{2} {\sT} 
         \frac{\int dx^{11} \sqrt{g_{1\!1 1\!1}} \delta \psi_{11}}
              { \int dx^{11} \sqrt{g_{1\!1 1\!1}}}.
\end{equation}
Note that this procedure allows for a nonlocal cancellation of the
vev of the auxiliary field in $d=4$. A condensate with equal
size and opposite sign at the observable wall could cancel the
effect and restore supersymmetry.
Using $\int dx^{11} \sqrt{g_{1\!1 1\!1}} \delta (x^{11})=1$, the 
auxiliary field is
found to be
\begin{equation}
  F_{\sT} ={\sT} \frac{1}{32 \pi (4 \pi)^{2/3}} 
           \frac{ g_8^2 \Lambda^3}{R_{11} M_{11}^3 }
\end{equation}
Similarly one can easily show that $F_{\sS}$ as well as the
vacuum energy vanish. This allows us then to unambiguously determine
the gravitino mass, which is related to the auxiliary field in
the following way:
\begin{equation}
  m_{3/2}=\frac{F_{\sT}}{{\sT}+{\sT}^*} 
 =\frac{1}{64\pi (4 \pi)^{2/3}} \frac{g_8^2 \Lambda^3}{R_{11} M_{11}^3}
 =\frac{\pi}{2} \frac{\Lambda^3}{M_{Planck}^2}.
\label{eq:gravitino-mass}
\end{equation}
As a nontrivial check one may calculate the gravitino mass 
in a different way. A term in the 
Lagrangian
\begin{equation}
  -\frac{\sqrt{2}}{192 \kappa^2}
   \int dx^{11} \sqrt{g} \bar \psi_I \Gamma^{IJKLMN} \psi_N 
                G_{JKLM},
\end{equation}
becomes the gravitino mass term when compactified to four dimensions.
Using the vevs of the $G_{IJK11}$ given by eq. (\ref{eq:vev-G}), one
can obtain the same result as eq.~(\ref{eq:gravitino-mass}). This is a
consistency check of our approach and the fact that the vacuum
energy vanishes in the given approximation.

It follows from eq.~(\ref{eq:gravitino-mass}), that the gravitino mass
tends to zero when the radius of the eleventh dimension goes to 
infinity. When the four--dimensional Planck scale is fixed to be 
the measured value, however, the gravitino mass in the strongly 
coupled case is expressed in a standard manner, similar to the weakly 
coupled case. To obtain the gravitino mass of the order of 
1 TeV, one has to adjust $\Lambda$ to be of the order of 
$10^{13}$ GeV when one constructs a realistic model by appropriately 
breaking the $E_8$ gauge group at the hidden wall.

In the minimization of the potential 
we have implicitly used the leading order approximation.
As was explained in a previous section, the next to leading order
correction gives the non--trivial dependence of the background metric
on $x^{11}$. Then the Einstein--Hilbert action in eleven dimensions
gives additional contribution to the scalar potential in the
four--dimensional effective theory, which shifts the vevs of the
$G_{IJKL}$. As a consequence, $F_S$ will no longer
vanish. 
Though this may be significant when we discuss soft masses,
it does not drastically change our estimate of the gravitino mass
(\ref{eq:gravitino-mass}) and our main conclusion drawn here is still
valid after the higher order corrections are taken into account.
In any case, these questions have to be addressed if one aims at
realistic models for particle physics.
%
%
\section{Summary and outlook}

In these lectures we have discussed the mechanism of gaugino 
condensation and flux stabilization within the heterotic scenario
in its simplest version. The picture could be easily generalized
to type II orientifolds or other types of string constructions,
leading to the notion of supersymmetry breakdown on a 
hidden brane.

There is, however, still a long way to go towards realistic model
building. One of the obstacles is the question of moduli stabilization
in string theory. Without a solution to this problem we shall usually
obtain so-called runaway vacua where e.g. coupling constants run to 
unrealistic values. Another obstacle is the appearance of
instabilities of the scalar potential once we include radiative 
corrections. We have discussed some aspects of this in section 6
when we included radiative (threshold) corrections to the
gauge coupling constants (45). If we would consider $f=S\pm\epsilon T$
and reinsert this into the $F-$terms in (32) we would obtain new 
contributions to the scalar potential leading to  minima with a negative vacuum
energy. This, of course, is nothing else than the problem of the
cosmological constant.

Recently there has been some revived interest in this discussion.
One aspect concerns the consideration of compactification of the
extra dimension on non-K\"ahler manifolds. 
Moduli are stabilized
with the help of fluxes of various antisymmetric tensor
fields. For more details and references see 
\cite{Becker:2003yv,Cardoso:2003sp,Curio:2001qi,Giddings:2001yu,Kachru:2002he}.
This allows the stabilization of many moduli already in the
supersymmetric framework in a rather general context and avoids 
cosmological moduli problems. 

Within the heterotic M-theory context there have been attempts
to go beyond the classical level \cite{Curio:2003ur}. In ref.
 \cite{Buchbinder:2003pi}
one finds a rather comprehensive discussion of moduli stabilization and
supersymmetry breakdown in a general set-up of heterotic M-theory
including 5-branes in the 
bulk. Moduli can be stabilized, but a 
large negative vacuum energy remains.
Similar results in the heterotic string theory have been reported
in \cite{Gukov:2003cy}. These results are, of course, also of
interest in the discussion of cosmological aspects of string
theory \cite{Kachru:2003aw}. So the mechanism of supersymmetry breakdown
through gaugino condensation still remains one the most
promising subjects in the discussion of realistic string model
building.

\section*{Acknowledgements}

We would like to thank the organizers, especially Gerard Smadja,
for their hospitality.
This work was partially supported by
the  European Community's Human Potential
Programme under contracts HPRN-CT-2000-00131 Quantum Space-time,
HPRN-CT-2000-00148 Physics Across the Present Energy Frontier and
HPRN-CT-2000-00152 Supersymmetry and the Early Universe.

%

%

\end{document}